\pdfoutput=1
 \documentclass[a4paper]{article}
\usepackage{amsmath}  % needed for \tfrac, \bmatrix, etc.
\usepackage{amsfonts} % needed for bold Greek, Fraktur, and blackboard bold
\usepackage{graphicx} % needed for figures
\usepackage[utf8]{inputenc}
\usepackage[affil-it]{authblk}
\usepackage[update,prepend]{epstopdf}
\usepackage{epsfig}
\begin{document}

\begin{center}
This article has been submitted to :

\textbf{American Journal of Physics}

Submission date: \textbf{August 5, 2014}
\end{center}

\title{Foucault precession manifested in a simple system}
% In a long title you can use \\ to force a line break at a certain location.

\author{Praveen S. Bharadhwaj}

\affil {Space Applications Centre, Indian Space Research Organization, Ahmedabad, India

(electronic mail: praveen@sac.isro.gov.in)}

\date{}

{\let\newpage\relax\maketitle} % title page is now complete

\begin{abstract}
This article aims to answer the question, ``What is the simplest system that embodies the essence of Foucault's pendulum?'' We study a very elementary idealized system that exhibits a precession behavior analogous to the classic pendulum. The system consists of a particle without inertial mass, constrained to an inclined plane that rotates about a vertical axis. Insights gained from this analysis are used to understand the rate of precession in a straightforward way.
\end{abstract}

\section{Introduction} % Section titles are automatically converted to all-caps.
% Section numbering is automatic.
Ever since Léon Foucault's first demonstration in Paris a century and a half ago, the pendulum's rate of precession with respect to local North has inspired a quest for elegant ways to explain it. `Foucault precession' is a slow rotation of the plane of oscillation of the pendulum. It is caused by the rotation of the earth, and its rate depends on geographical latitude. This classic experiment is a perennial favorite at science museums and similar venues, apart from its value in  teaching certain ideas in mathematical physics.

The precession effect has been analyzed with reference to a variety of concepts, including Coriolis forces, parallel transport, geometric phase, the Gauss-Bonnet theorem, the construction of a locally tangential conical surface with an embedded coordinate system, and the division of a latitude circle into a series of infinitesimal great-circle segments.  The diversity of this mathematical arsenal reflects the phenomenon's enduring richness as a subject of study.\cite{jordan-maps, bergmann, wiley-paradox, Krivor, boersma-mathician, oprea-geometry, wiley-paradox, hart-miller-mills-geometric, stanovik_elastic} One interesting strand of this work is the devising of mechanical and virtual artifacts that show analogous behavior and help to illuminate the subtleties of the problem.\cite{weltner-model, cleonis, gil-device, sears-model, santander_mech_chariot}

This article aims to answer the question, ``What is the simplest system that embodies the essence of Foucault's pendulum?''

At first sight it may seem that our model would involve a particle with \textit{mass and inertia}, a \textit{centrally directed force}, and the \textit{transportation} of the whole system over a curved surface. However, we shall show that a much simpler model, lacking any of these elements, serves as an interesting mathematical parallel to the original Foucault pendulum. This connection will be used to understand the precession of the pendulum itself in a straightforward way. We first explore this route graphically by presenting the results of numerical simulations. Next we discuss the underlying physical and mathematical principles.

\section{An inertia-less particle \textit{without} a centrally directed force}
Consider a massless (inertia-free) particle that is constrained to remain in contact with a certain planar surface that we will call the constraint plane. There is no gravitation, no centrally directed force, and no friction. The constraint plane is inclined at a fixed angle with respect to the XY plane, and it rotates at a constant clockwise angular velocity about the Z axis. The particle is initially at some distance from the origin. 

Our constraint that the particle must remain on the plane is specified more carefully as follows: If the particle should find itself momentarily displaced from the plane, it will move very rapidly along the normal and towards the nearest point on the plane, until it is once again in contact. If this process is fairly rapid compared with the rotation of the constraint plane, then the particle is effectively bound to the plane. 

It is perhaps hard to think of a system simpler than an inertia-less, frictionless particle that is mindlessly chasing the nearest point on a slowly rotating inclined plane. Yet this problem is mathematically akin to Foucault's pendulum, and it follows the well known relation between latitude and rate of precession. The tilt angle of the rotating constraint plane in our model takes the place of the complement of latitude in the original Foucault problem; the locus of the particle within the plane is equivalent to the precession of the pendulum's extrema. 

\begin{figure}[h!]
\centering
\includegraphics[ width=4in, angle=-90]{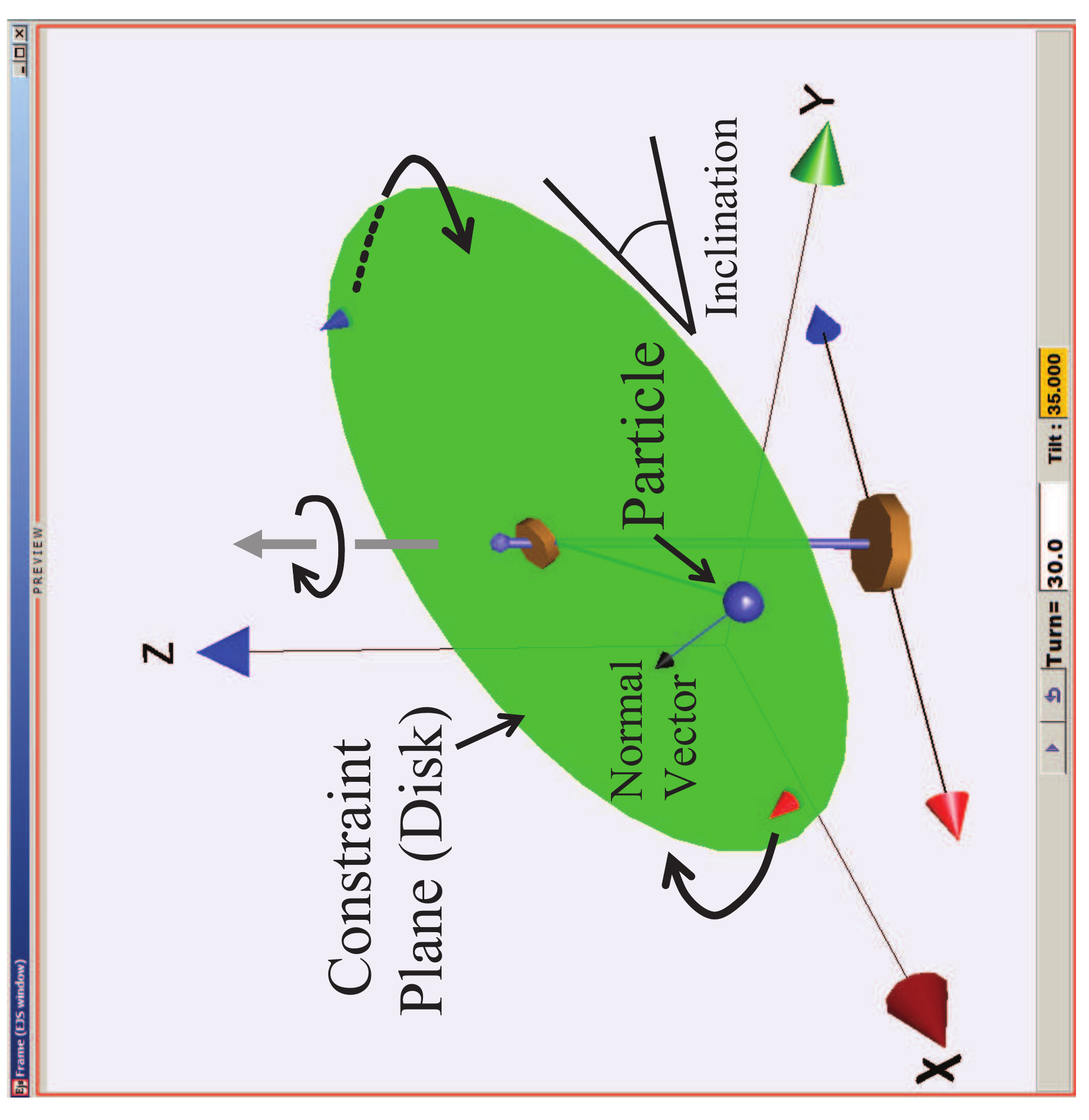}

\caption{Massless particle on a constraint plane. Animation view in \textit{Easy Java Simulations (EJS)}.}
\label{basicdisk}
\end{figure}

We will discuss the results of numerical simulations on the computer program \textit{Easy Java Simulations} (\textit{EJS}), which can be used to solve ordinary differential equations (ODEs) and to create animations of the evolving system on a graphical canvas\cite{ejsHome,ejs_compadre}.

Fig.~\ref{basicdisk} 
shows the 3-D (perspective) animation panel of our \textit{EJS} simulation, in which the disk represents the constraint plane, and the sphere represents a massless particle that is constrained to remain on the plane. (In the rest of this article we will refer interchangeably to \textit{particle} and \textit{ball}; and likewise, to \textit{constraint plane} and \textit{disk}). 

Recall that there is no gravitation; the ball does not roll ``downhill''. If the disk were stationary, the ball too would be stationary. Without inertial mass, the particle cannot gather momentum by integrating force over time. It does \textit{nothing but} follow very closely the nearest point on the constraint plane.
\section{Numerical simulations}\label{ejs_results}
\subsection{Constraint plane with zero angle of inclination}
We start with the trivial case where the disk is parallel to the XY plane. As expected, the disk can convey no motion to the particle, since the velocity of the elemental contact area is directed at right angles to the normal vector to the plane. With no friction present, the ball remains at its initial position forever as the surface merely slips past with no effect. We can also say that the particle is constantly slipping \textit{back} with respect to the disk because it is able to satisfy our constraint without having to move.
\subsection{Constraint plane inclined at 10$^{\circ}$}\label{subsec_10_degs}

\begin{figure}[h!]
\includegraphics[trim=50 250 12 0, clip, width=\textwidth, angle=0]{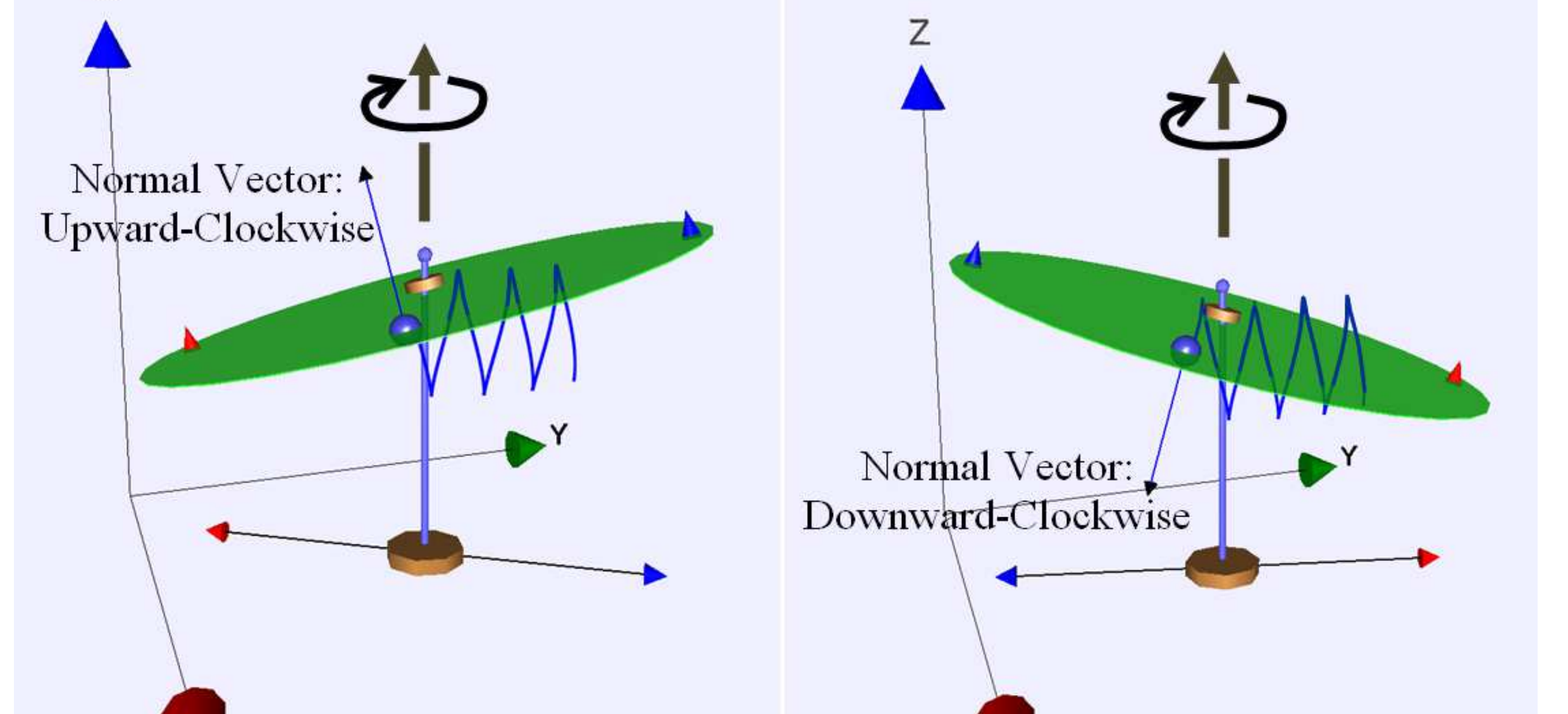}
\caption{With the constraint plane tilted at 10 degrees, 
the particle executes a bobbing walk around the center.}
\label{10_degs}
\end{figure}

Figure~\ref{10_degs} shows the system with the disk inclined at 10 degrees from the XY plane while rotating about a vertical axis. As before, the ball lags appreciably with respect to the disk, but now it also performs an up-and-down bobbing motion as it tracks the nearest point on the disk.
Most importantly, we observe that the vertical oscillation has a clockwise drift or “walk” superimposed on it. Visually, it suggests a small friction-like force dragging the ball along the surface, but the true cause is as follows: The nonzero tilt angle now means that the normal vector (i.e. the direction of pursuit) will, at times, have a small component along the horizontal-clockwise direction. As the disk rotates, the particle’s velocity vector will cycle through the upward clockwise and downward-clockwise directions as well as the purely upward and purely downward directions. 
This process adds up to a 360$^{\circ}$ walk during the course of several revolutions of the disk. 

Since the elemental velocity of the disk is always directed along a tangent to a horizontal circle, no radial displacement can be induced in the particle. It remains at the initial distance $R$ from the origin. Its locus, however complex it may turn out to be, will conform to a sphere of radius $R$. (See also Fig.~\ref{60_degs})

\subsection{Disk inclined at 90$^{\circ}$}
Considering now a tilt of 90$^{\circ}$, the contact point velocity is directed exactly along the normal to the disk, allowing no slip or leeway between the ball and the disk. This direction is always tangential to a horizontal circle that contains the initial location. The particle is swept bodily along this circle at the same angular velocity as the disk itself. (If the particle had possessed inertial mass, it would have been slung outwards along a radial path on the disk; see Appendix).
\subsection{Inclination of 60$^{\circ}$ and 88$^{\circ}$}
\begin{figure}
\centering
\includegraphics[ width=\textwidth, angle=0]{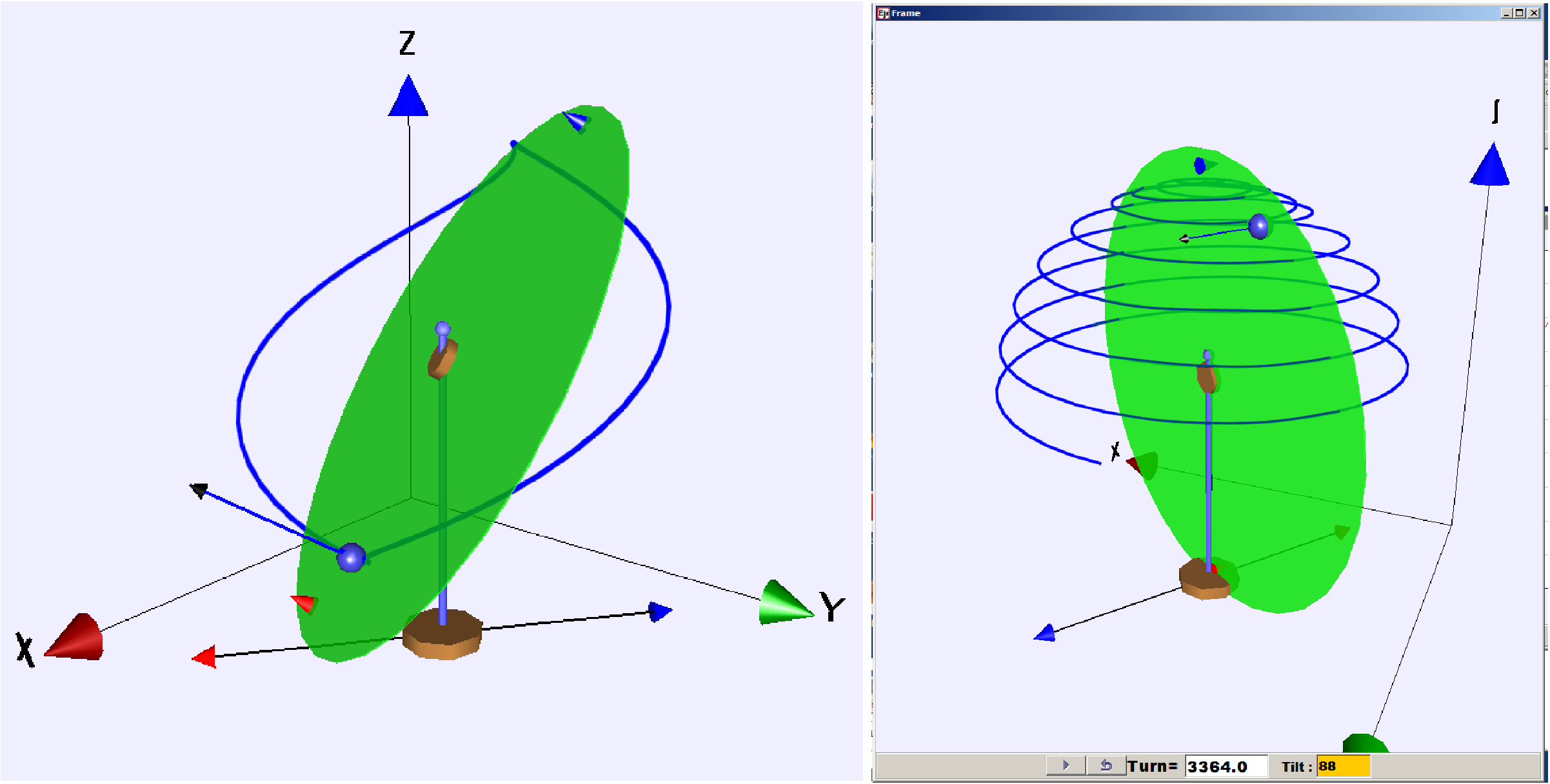}
\caption{(a) Simulation with constraint plane at 60 degrees. The particle is swept around one circuit during exactly two revolutions of the disk. (b) Constraint plane at 88 degrees.}
\label{60_degs}
\end{figure}
With the inclination angle set at 60$^{\circ}$, the vertical oscillation is much larger than in Case-B. The locus has only two cusps. The clockwise component is large enough to whisk the particle through one turn about the Z~axis, during two turns of the constraint plane (Fig.~\ref{60_degs}). This corresponds to a relative~precession~rate of 50\% with respect to the disk’s reference frame – a result that matches the classic Foucault pendulum at a latitude of (90 – 60) = 30$^{\circ}$. Also shown in Fig.~\ref{60_degs} is an interesting locus with the disk inclined at 88$^{\circ}$. The ball walks about a quarter of the way around the disk during seven or eight turns of the latter. As noted in Sec.~\ref{subsec_10_degs}, the locus always conforms to a sphere.
\section{Particle motion with respect to the constraint plane}
A rotational transformation is used in the simulation to compute the particle's locus in the disk's rotating coordinate frame. Despite the complex nature of the trace in our fixed coordinate system, the particle moves uniformly along an anticlockwise circle on the disk, except for the case of 90 degree tilt -- where the particle is stationary.
\begin{figure}[h]
\centering
\includegraphics[trim=8 0 0 20, width=2.5in, angle=-90]{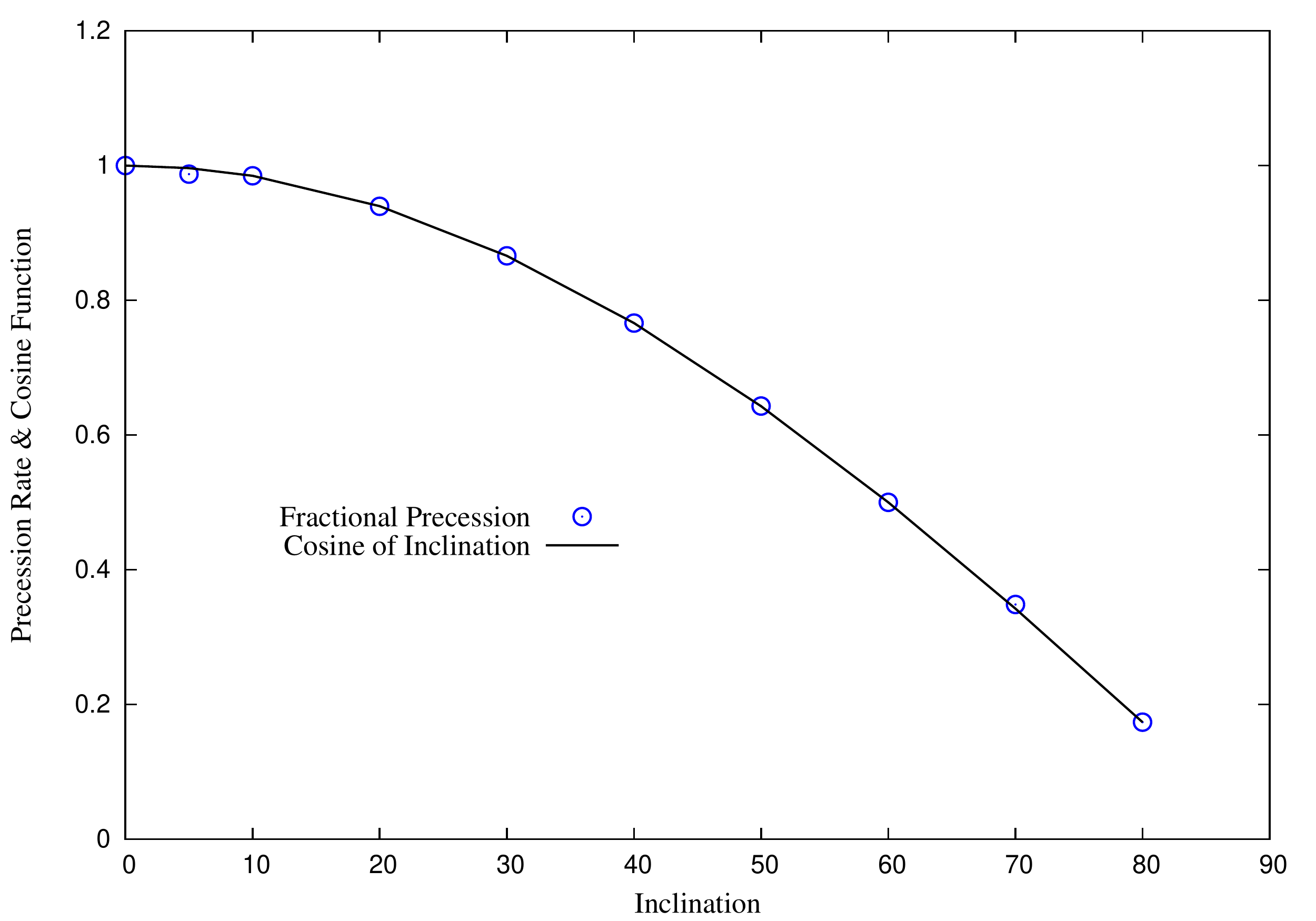}
\caption{Fractional rate of precession, plotted as a function of the angle of inclination and compared with the cosine.}
\label{precession_plot}
\end{figure}
Fig.~\ref{precession_plot} shows the simulated fractional rate of precession as a function of the angle of inclination.  This is the angular velocity of the particle with respect to the disk, divided by the angular velocity of the disk in our “absolute” frame. Its agreement with the cosine law shows a similarity of behavior with the Foucault pendulum, which precesses at a daily rate that is proportional to the cosine of the inclination of the local horizontal with respect to the earth’s axis. 

The motion of the particle within the reference frame of the disk can be deduced from the following fact:
\begin{quote}\textit{All elemental areas of an inclined rotating disk, lying at an equal distance R from the centre, have the same in-plane component of linear velocity, which is given by} $R\omega\cos\phi$ \textit{where $\omega$ and $\phi$ are the angular velocity and inclination of the disk.}\end{quote}

This is clear enough if we project the angular velocity vector $\vec \omega $  onto the normal axis of the disk to get $\omega_D\equiv\left|\vec\omega_D\right|=\left|\vec\omega\right|\cos\phi$  , i.e. the component of angular velocity around the disk’s own axis. The in-plane velocity of any element at a distance \textit{R} from the center is then $R\omega_D=R\omega\cos\phi$. 
\begin{figure}[h]
\centering
\includegraphics[trim=400 200 0 20, width=3.in, angle=0] {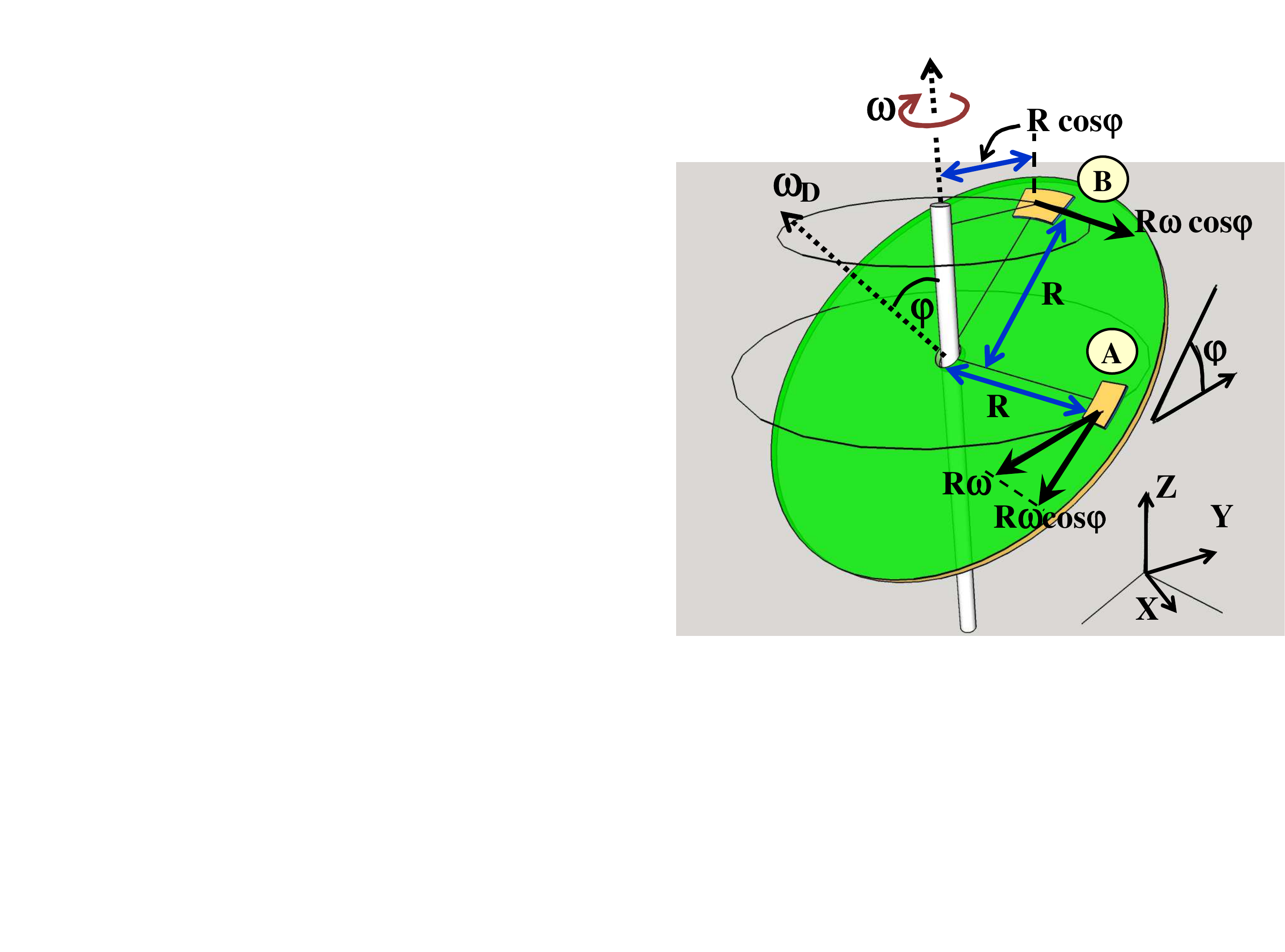}
\caption{Two elemental areas equidistant from the center of the disk have the same in-plane component of linear velocity.}
\label{intuitive}
\end{figure}

It may be instructive to distil the intuitive spirit of this result by considering two special elemental areas whose in-plane linear velocities can be compared by mere inspection of Fig.~\ref{intuitive}. Element A lies on the axis of inclination while element B lies on a line perpendicular to it. As we increase the angle of inclination from zero to 90$^{\circ}$, the linear velocities of the two elements are affected in the following ways: Element A remains at a distance $R$ from the axis, and has a linear velocity $R\omega$ that is not affected by the inclination. However, this vector (being always horizontal) swings out of the inclined plane and subtends an angle $\phi$. The in-plane component is then $R\omega\cos\phi$. The same tilt $\phi$ causes Element B to approach the axis of rotation, so that the element has a reduced linear velocity $R\omega\cos\phi$. This vector is directed  along the surface, so that the in-plane component\cite{student exercise} is still $R\omega\cos\phi$.

If we now consider a particle in frictionless contact with (or close proximity to) the disk, the lateral slip of the surface with respect to the particle is given by $R\omega\cos\phi$. The velocity of the particle with respect to the disk is therefore $-R\omega\cos\phi$. We have seen in Section~\ref{subsec_10_degs} that $R$ is constant, because the radial velocity is always zero. The locus of the particle over the disk is therefore a circle of radius $R$ traced at a constant  angular  velocity $-(\omega\cos\phi)$. 

This is the precession one would observe ``on board'' the rotating disk. 

\section{Inrtia-free particle-on-disk \textit{versus} Foucault's pendulum: a shared underlying principle?}
We have noted a behavioral similarity between our simulated model and the Foucault pendulum. We now try to identify a common principle from two different, but interlinked, perspectives.
\begin{figure}[h]
\centering
\includegraphics[trim=100 475 100 25, width=3.75in, angle=0] {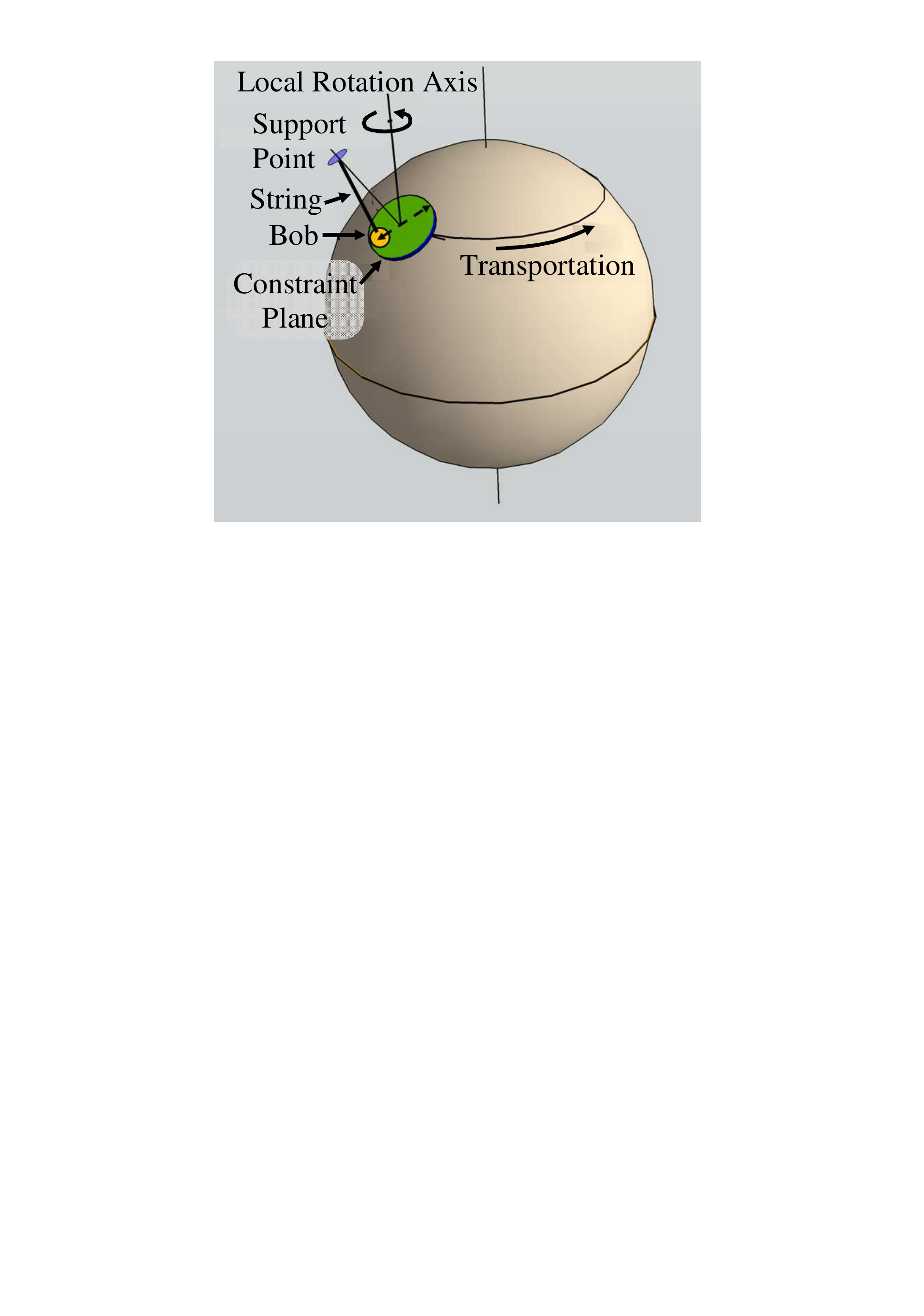}
\caption{Physical pendulum's constraint plane with local rotation and transportation}
\label{globedisk}
\end{figure}
A real pendulum is constrained to remain on a spherical surface of fixed radius equal to the length of the string. For small amplitudes this can be approximated by a constraint plane whose inclination with respect to the earth's axis depends on latitude. The effect of the earth's rotation can be decomposed into a rotation about a local axis and a transportation along the latitude circle (Fig.~\ref{globedisk}). In what follows, we neglect the transportation, and concentrate on the rotation about a local axis passing through the bob's center of oscillation (equilibrium position). We consider the rotation to be occurring about this axis, which we treat as being fixed in space.

\subsection{The pendulum's locus considered as an inertia-less mathematical object}
If the system were not rotating, the pendulum would periodically retrace a bounded path that would be fixed with respect to the distant stars. This locus, considered as a mathematical object, is inertia-free in the following sense: if an external influence has recently been causing the locus to drift, then the withdrawal of that influence will not be followed by any \textit{further} drift of the locus. The integrated history of any previous external effects may possibly be reflected in the present amplitude  and present orientation of the path, but not as a continued motion of the path \textit{per se} after the external influence has ceased.

The constraint plane's rotation dictates a continuous infinitesimal realignment of the pendulum's locus in order to remain on the plane. Since the pendulum's arm or string is always normal to the constraint surface, it can only communicate a displacement normal to the bob's motion. The continuous infinitesimal realignment of the locus with the moving constraint plane involves nothing but a close pursuit of the surface along a normal. 

To sum up, the bob's locus behaves as a set of bounded, persisting, inertia-free points that are constantly trying to follow the constraint plane.
The Foucault pendulum's locus (but \textit{not} the bob itself) shares three defining attributes with our simple system:
\begin{enumerate}
\item Absence of inertia in the sense explained above (zero order behavior)
\item Absence of in-plane perturbations
\item A compelling need to pursue the constraint plane closely along the normal
\end{enumerate}
These points of similarity lead to a mathematical likeness between the two systems. The constraint disk slips past the locus of oscillation in much the same way that it slips past the massless particle. It also impels the massless particle and the pendulum's locus along a normal in both systems. The tangential slip and the normal push-pull interaction are mutually decoupled processes that can be analyzed separately, precisely because there is no inertia. In the Appendix, we probe deeper into this notion of an ``inertia-less'' locus, and compare its behavior to a particle \textit{with} inertia that is constrained to a revolving inclined disk.

\subsection{Precession in terms of ``real'' forces}
Up to this point we have refrained from talking about a ``force'' acting on the particle, since this would not be meaningful for a massless object. In a real pendulum, an approximately planar constraint surface is established by the position of the support, length of the string and direction of gravitational force. In practice, the string would have a finite elastic constant. We discuss here the action of the physical forces that actually create our somewhat abstract constraint plane\cite{stanovik_discuss}.

Consider an abrupt, infinitesimal rotation of the pendulum's entire system around the local Z axis as defined in Fig.~\ref{globedisk}. (As stated before, we are ignoring transportation over the earth's surface). This would result in a configuration where the old position of the bob does not necessarily lie on the new, post-rotation constraint plane. The string would momentarily stretch or contract by an infinitesimal length, which would change the elastic force exerted by it. The force balance would alter minutely, so that the bob would be drawn towards the new constraint plane along the direction of the string.

If we now consider a series of such incremental rotations, the successive displacements would integrate into the now familiar bobbing walk around the local Z axis, superimposed on the pendulum's natural radial oscillation (see Appendix). If the string's elastic constant is made arbitrarily stiff, we approach an ideal case that we can analyze through our favorite mathematical formalism.

\section{Does the constraint plane really rotate?}
We must now confess to, and clear up, a piece of non rigorous hand-waving that we have been indulging in. It is clear enough how physical forces operate to define the constraint plane, but is this plane really endowed with a polar grid that rotates with it, like lines engraved on a solid disk? We do know from geometrical reasoning how a large compass rose, laid flat on the earth's surface, orients itself with respect to the stars as a function of time (Fig.~\ref{globedisk}). We recognize that it remains always parallel to the pendulum’s constraint plane. We have therefore na\"{\i}vely (but conveniently) transferred the rose’s definition of North onto our moving plane. This step, logically dubious at best, has led us on a shortcut to some useful information about precession.

\begin{figure}[!h]
\centering
\includegraphics[trim=20 375 80 30, width=4in, angle=0] {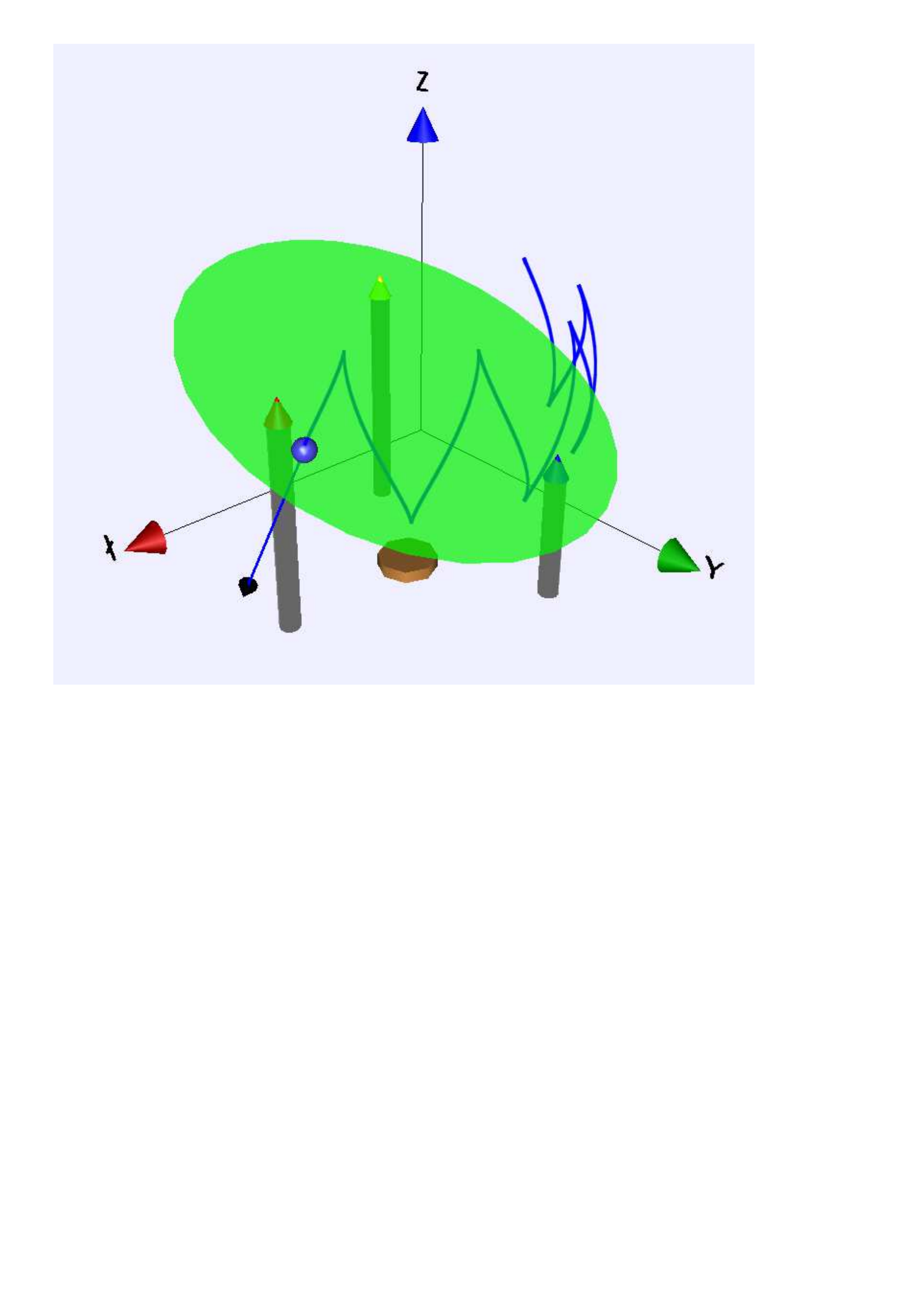}
\caption{An arrangement without explicit rotation, physically equivalent to that described in Sec.~II. Three linear actuators cause the disk to wobble in a manner that parallels the rotating system.}
\label{wobbler}
\end{figure}To judge only by its effect on the particle, Fig.~\ref{wobbler} is equivalent to our original system. The disk is ``wobbled'' by three linear actuators that are coordinated so that it is always parallel to a hypothetical disk that is indeed rotating. With respect to the laboratory frame, the particle in this system will describe a locus identical to the one in the rotating-disk system. The rotation that we cheerfully ascribed to the original disk had no physical contribution in this sense.

That said, nothing prevents us from mounting a finely crafted, motorized compass rose on top of the disk in Fig.~\ref{wobbler}. Nothing prevents us from running the motor, and turning the rose, at any convenient speed that would help to deduce something interesting about Foucault’s pendulum. The constraint plane does all the pushing and pulling; the compass rose does the book-keeping. 

One could even maximize pedagogical verisimilitude by arranging to transport the whole apparatus over the surface of a colossal globe (trundling along, for example, atop an ancient Chinese south-seeking chariot\cite{bergmann,santander_mech_chariot,gil-device}).

\section{Conclusion}
We have described a simple idealized system that exhibits the same precession behavior as Foucault's pendulum. The model consists of an inertia-less particle that is constrained to remain on an inclined rotating plane. The rate of precession can be derived from basic trigonometry, without calling up the heavy artillery of any mathematical formalism. The underlying equivalence of this system to the Foucault pendulum is evident if one recognizes the locus of the pendulum to be an inertia-less zero-order entity that is driven by an inclined rotating constraint plane with no tangential interaction.

Being bounded and periodic, the pendulum's locus behaves as a persisting object in its own right, with a definite state at any instant. The state evolves according to zero order dynamics with the constraint plane's rotation as the driving input\cite{perturb}. Precisely because there is no inertia, the tangential and normal evolutions can be decoupled and understood separately. Our basic particle-on-disk model is perhaps the simplest example of such behavior.
\appendix  
\section{On the ``inertia-less-ness'' of the pendulum's locus; Or, a plague of locus'}
\begin{figure}[h!]
\centering
\includegraphics[trim=100 340 30 160, width=3.5in, angle=0] {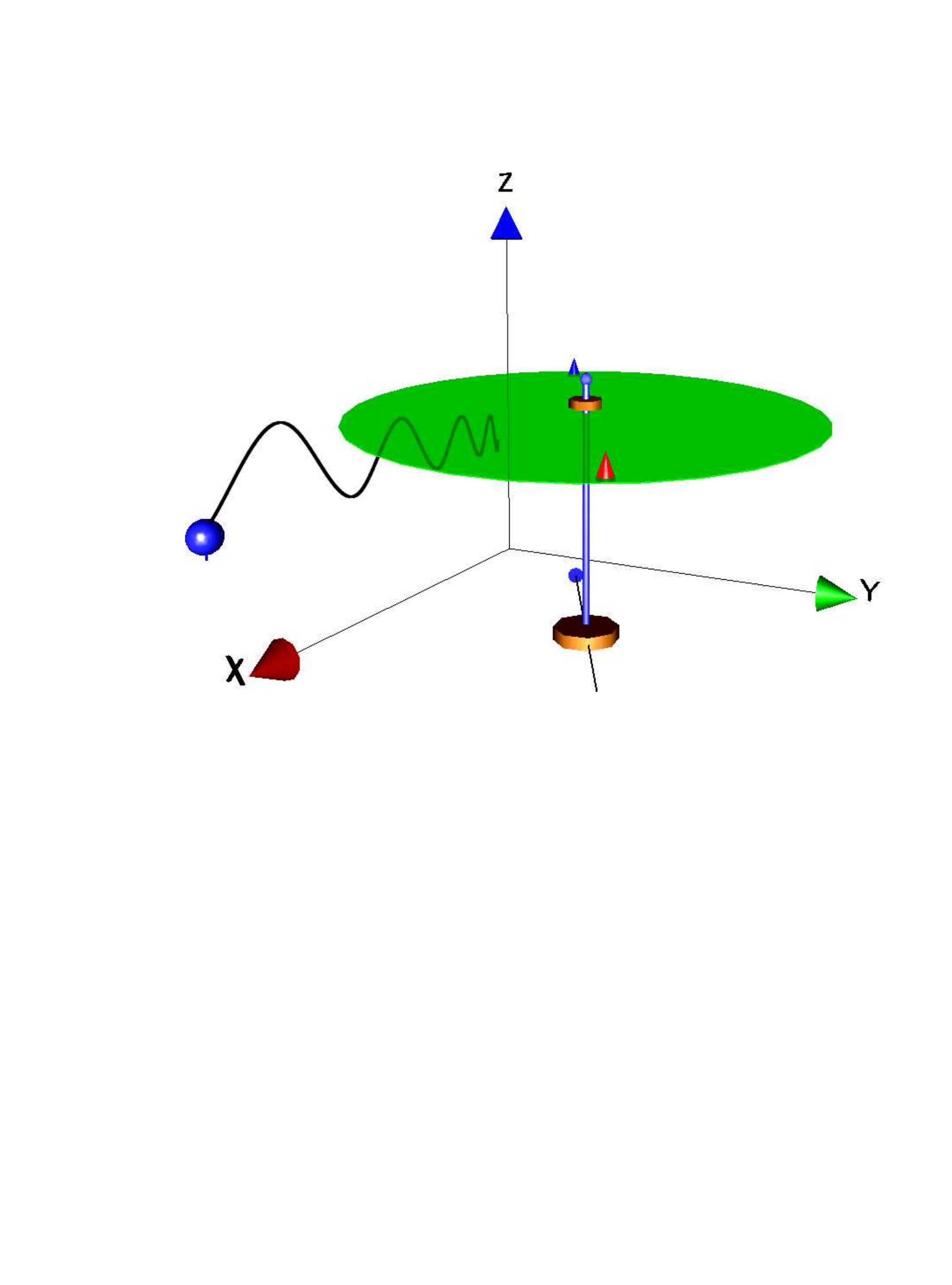}
\caption{When inertial mass is added, the particle gathers momentum and travels away from the system - even when the inclination is only 5$^{\circ}$.}
\label{with_inertia}
\end{figure}

What happens to the ball-on-disk if we add some inertial mass? Fig.~\ref{with_inertia} is a simulation of that system. 
The particle builds up momentum by integrating force, and travels outwards with ever increasing in-plane velocity. The locus is unbounded and nonperiodic. (Even if the disk were to stop rotating at some point, the particle would coast along forever, which is something that our massless particle would not do).

\begin{figure}[h]
\centering
\includegraphics[trim=100 289 100 0, width=4.5in, angle=0] {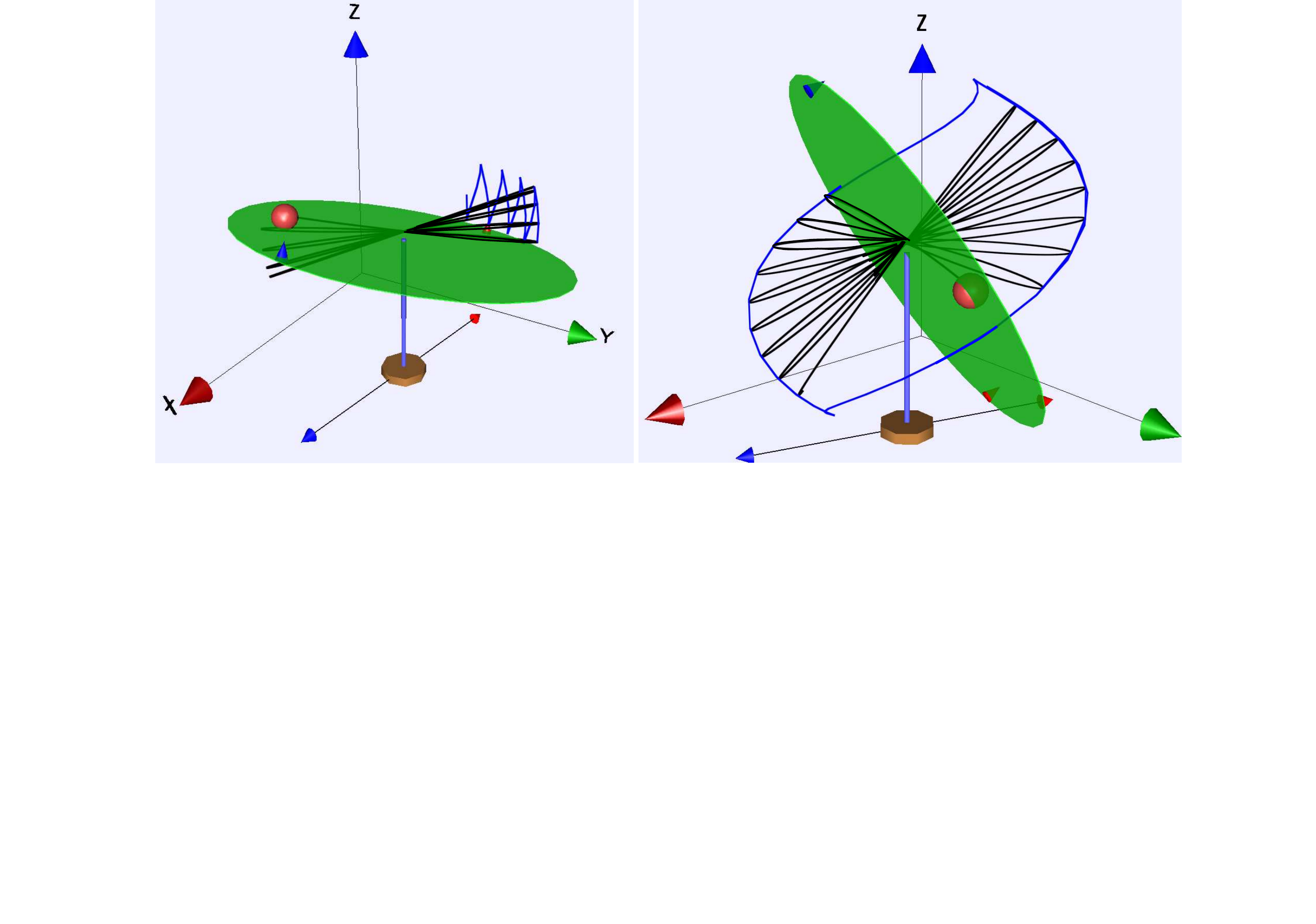}
\caption{A restoring force has been added to the massy particle, converting it into an oscillator. Radial trace: Recent locus of bob. Peripheral trace: Locus of an extreme point of swing (``locus of the locus''). These simulations were run with inclinations of (a) 10$^{\circ}$ and (b) 60$^{\circ}$.}
\label{pendulum}
\end{figure}

We can prevent the particle from escaping by binding it to the general vicinity of the origin with a centrally directed linear force. This, of course, makes it a simple harmonic oscillator. Furthermore, it is a Foucault oscillator; it precesses exactly like the classic pendulum. Fig.~\ref{pendulum} shows the locus of the pendulum over the last few cycles of oscillation for inclination angles of 10$^{\circ}$ and 60$^{\circ}$. Also shown is the locus of an extreme point over a longer time interval – what we could call the “locus of the locus” or even the \textit{meta-locus}. This latter trace corresponds, of course, to that of the inertia-less particle that we started with (\textit{cf} Figs.~\ref{10_degs} and \ref{60_degs}a).

Being bounded and periodic, the locus of the bob can be regarded as an object in its own right, with a definite state at any instant. The state evolves according to inertialess, or zero order, dynamics, with the moving constraint plane as a driving input. It is the absence of inertia that permits a convenient decoupling of the normal and tangential evolutions -- and helps our analysis to fit on the back of an envelope. Our basic ball-on-disk model is perhaps the simplest example of such behavior.


\begin{thebibliography}{99}
\bibitem{student exercise}Students may find it interesting to generalize this result to any point on the disk, without recourse to the projection of angular velocity.
\bibitem{perturb}This general line of argument can perhaps be described as a heuristic form of perturbation theory. Starting with the known properties of the locus when the constraint plane is stationary, we extend our understanding to include a rotational perturbation.
\end{thebibliography}
\end{document}